\documentclass[english,prl,aps,reprint,longbibliography,superscriptaddress]{revtex4-2}
\usepackage[T1]{fontenc}
\usepackage{amssymb}

\makeatletter
\usepackage{babel}

\makeatother

\usepackage{babel}
\begin{document}
\title{Absence of Parity Anomaly in Massive Dirac Fermions on a Lattice}
\author{Shun-Qing Shen}
\email{sshen@hku.hk}

\affiliation{Department of Physics and State Key Laboratory of Optical Quantum Materials, 
The University of Hong Kong, Pokfulam Road,
Hong Kong, China}

\begin{abstract}
The parity anomaly for Dirac fermions in two spatial dimensions has shaped perspectives in quantum field theory 
and condensed matter physics. In condensed matter it has evolved as a mechanism 
for half-quantized Hall responses in systems described by massive Dirac fermions. 
Here we reexamine the issue on a lattice and show that the half-quantized Hall conductivity is absent 
for massive Dirac fermions when lattice regularization is properly implemented and 
the translational invariant symmetry is taken into account. We realize that a single massive Dirac cone 
on a lattice always leads to an integer quantized Hall conductivity and to the half-quantized Hall 
conductivity only in the unphysical limit of infinite momentum cut-off. 
The half-quantized Hall conductivity appears with nonzero longitudinal conductance
as a signature of a single massless Dirac cone on a lattice. Consequently, 
the parity anomaly is a property of massless Dirac fermions in a semimetal/metal, 
not of massive Dirac fermions in an insulator on a lattice.

\end{abstract}
\date{October 31, 2025}
\maketitle

\paragraph*{Introduction-} When the massless Dirac fermions move in an external field 
$A_{\mu}$,
which is minimally coupled to the fermions in the Dirac equation,
\begin{equation}
\gamma^{\mu}\left(i\partial_{\mu}-eA_{\mu}\right)\psi=0,
\end{equation}
a nontrivial vacuum current is induced \cite{Jackiw-84prd,Avareme-84npbz,Niemi-83prl,Redlich-84prl},
\begin{equation}
\left\langle j^{\mu}\right\rangle =\pm\frac{e^{2}}{2h}\epsilon^{\mu\nu}E_{\nu}\label{eq:vacuum-current}
\end{equation}
where $E_{\nu}$ is an electric field. The sign ambiguity arises from
the necessary regularization procedure to avoid the divergence of
the vacuum polarization. This phenomenon is named parity anomaly.
In 1983, Niemi and Semenoff first introduced
a mass term $m$ in Eq. (1), which breaks the time-reversal symmetry,
and found that 
$\left\langle j^{\mu}\right\rangle =\mathrm{sgn}(m)\frac{e^{2}}{2h}\epsilon^{\mu\nu}E_{\nu}$
\cite{Niemi-83prl}, which is only determined by the sign of mass. When the mass term tends to vanish, $m\rightarrow0$,
the current still survives. They
thought that this approach solves the sign problem in parity anomaly.
At the same time, Redlich introduced a heavy Pauli-Villars regulator
field $\lim_{M\rightarrow\infty}I_{eff}[A,M]$ and subtracted from
the action $I_{eff}[A]$ for the massless Dirac fermions coupled to
the gauge field, which regulates the ultraviolet divergence in the
calculation of $I_{eff}[A]$ \cite{Redlich-84prl}. He found that $\lim_{M\rightarrow\infty}I_{eff}[A,M]$
contains a parity-nonconserving topological term, which may determine
the sign of the vacuum current in Eq. (\ref{eq:vacuum-current}). These
two approaches are actually distinct, Niemi-Semenoff approach is
based on the massive Dirac fermions by taking the limit of $m\rightarrow0$
finally while Redlich approach still keeps the picture of massless
Dirac fermions by introducing an additional and infinite mass term.

Soon after Semenoff \cite{Semenoff-84prl} applied the theory to electrons on a honeycomb
lattice with different site potentials $+M$ and $-M$ on A and B sublattice 
sites, which hosts two massive Dirac fermions with opposite signs
of mass $+M$ and $-M$ at the points or valleys $K$ and $K'$. He claimed that
the Hall conductivities for the two massive Dirac fermions are 
$\sigma_{\pm}=\pm{\rm sgn}(M)\frac{e^{2}}{2h}$,
respectively. Although the total Hall conductivity as the sum of two
Hall conductivities for the massive Dirac fermions is equal to zero,
$\sigma_{total}=\sigma_{+}+\sigma_{-}=0$, 
the difference $\sigma_{+}-\sigma_{-}={\rm sgn(M)\frac{e^{2}}{h}}$. He interpreted
that an electric field can induce a valley Hall current, which is the difference of
the currents associated with the two valleys, 
$\mathbf{j}_{v}=\mathbf{j}_{+}-\mathbf{j}_{-}={\rm sgn}(M)\frac{e^{2}}{h}\hat{z}\times\mathbf{E}$, and
proposed that this could lead to the realization of parity anomaly
in condensed matter. 
The effect was named the quantum valley Hall effect after the discovery of the graphene, 
and extensively studied theoretically and experimentally \cite{Gorbachev-14science,Lensky-15prl,
Ju-15nature,Sui-15natphys, Shimazaki-15natphys,Ren-16RPP, KazanTsev-24prl}. In 1988, Haldane \cite{Haldane-88prl} 
proposed that the presence of periodic magnetic flux $\phi$ in the honeycomb lattice
can manipulate the amplitude and even sign of masses of the two Dirac
fermions. When two masses have the same sign, and the total Hall conductivity
is $\mathrm{sgn}(M)\frac{e^{2}}{h}$. This gives rise to the quantum
Hall effect without Landau levels in two dimensions,
which has been confirmed experimentally \cite{Chang-13science,Checkelsky-14natphys,Kou-14prl,Yu-10science, Chu-11prb, Qiao-10prb}. 
Thus Semenoff's proposal leads to a picture of the parity anomaly for massive Dirac fermions on a
lattice \cite{Qi-11rmp, Tokura-19nrp, Chang-23rmp}, which has extensive impact in the field of topological insulator
and 2D materials over the past forty years.

In this article, we find that the parity anomaly for massive Dirac
fermions on a lattice is actually absent. 
The half quantized Hall conductivity of massive Dirac fermions is only valid in the
continuous case when the cut-off of the wave vector is taken to be
infinity. On a lattice, the first Brillouin zone is always finite and periodic, 
which leads to the strict constraints to realize massless
and massive Dirac fermions on a lattice. The half quantized Hall conductivity actually 
reflects the parity anomaly for massless Dirac fermions instead of massive Dirac fermions. 
Thus some theories related to the parity anomaly for massive Dirac fermions on a lattice, 
such as the quantum valley Hall effect in 2D materials metamaterials, the half quantized Hall conductivity of the gapped surface
states and axion insultor, need to be reassessed carefully.

\paragraph*{On half quantized Hall conductivity of massive Dirac fermions-}

By convention of condensed matter physics, we use the Hamiltonian
formalism to study the physics of the massive Dirac fermions on a
translationally invariant lattice. Two-dimensional massive Dirac
fermions in Semenoff's proposal in a continuous form reads 
\begin{equation}
H=v\hbar k_{x}\sigma_{x}+v\hbar k_{y}\sigma_{y}+mv^{2}\sigma_{z}.\label{eq:massive-Dirac}
\end{equation}
where $m$ is the effective mass, $v$ is the effective velocity and
$\sigma_{i}$ are the Pauli matrices. The mass term breaks the time
reversal symmetry explicitly. By means of the Kubo formula for an
external response, the Hall conductivity can be evaluated by integrating
the Berry curvature over momentum space \cite{Xiao-10rmp}. When the chemical potential
is located with the band gap between the valence and conduction band,
the Hall conductivity is given by 
\begin{equation}
\sigma_{H}=\frac{e^{2}}{2h}\left [\mathrm{sgn}(m)-\frac{mv^{2}}{\sqrt{(v\hbar k_{c})^{2}+m^{2}v^{4}}}\right ]\label{eq:massive-Hall-conductivity}
\end{equation}
where $k_{c}$ is the cut-off of the wave vector. 
The Hall conductivity is not half quantized for a finite $k_{c}$, and approaches to 
$\sigma_{H}={\rm sgn}(m)\frac{e^{2}}{2h}\label{eq:m-Hall-conductivity}$
only if $k_{c}$ is take to be infinity. The conductviity survives even in the limit of $m\rightarrow 0$.
Here we reproduce the result obtained by Niemi and Semenoff in the quantum field theory \cite{Niemi-83prl}.

When the valence band of the negative energy is fully filled, the
system is insulating because of the presence of the energy gap $2mv^{2}$. 
In this case, how can electrons move to response to an external field? 
For an insulating and noninteracting electron system, it is known that the Hall conductivity
must be an integer, which corresponds to the number of chiral edge states around boundary \cite{Hatsugai-93prl}. 
As the number of the edge states is always an integer,
an serious issue arises:\textit{how can an insulating system give rise to the half quantized Hall effect when all electrons
are localized?}
 
\paragraph*{Periodicity of the Brillouin zone and integer quantized Hall conductivity-}

On a translationally invariant lattice, the first Brillouin zone in
the reciprocal lattice  space is always finite and periodic, which 
reflects the compactness of the reciprocal lattice
space. Consequently, it leads to the strict constraints to
realize single massless and massive Dirac fermions on a lattice. For simplicity,
we consider Eq. (\ref{eq:massive-Dirac}) on a square lattice. For $k_{y}=0$,
the spin polarization of the electron states is determined by the
value of $k_{x}$. In the conduction band, at $k_{x}=0$, the electron
is polarized is along the z-direction, and its direction is determined
by the sign of $m$. With increasing $k_{x}$, the linear term becomes dominant, and the electron is intended
to be polarized along the x-axis, and the directions at $k_{x}$ and
$-k_{x}$ are almost opposite for $v\hbar k_{x}\gg mv^{2}$. On the other hand, 
due to the periodicity of the Brillouin zone, the electron state $\left|-k_{x}\right\rangle $
is equivalent to $\left|-k_{x}+K\right\rangle $ where $K=2\pi/a$
(a is the lattice space). Thus at $k_{x}=+K/2$ and $k_{x}=-K/2$,
the two states are actually identical, and the spin polarization must
be along the same direction. This contradicts with the fact that the
opposite spin polarizations in $\left|\pm k_{x}\right\rangle $ in
Eq. (\ref{eq:massive-Dirac}) are opposite. So Eq. (\ref{eq:massive-Dirac})
is only valid for a small $k$. Thus additional modification is necessary
to make the model valid in the whole Brillouin zone.

To solve the contradiction, one may introduce a k-dependent quadratic
correction to the mass term in Eq. (\ref{eq:massive-Dirac}), $mv^{2}\rightarrow m(k)=mv^{2}-Bk^{2}$
where $k^{2}=k_{x}^{2}+k_{y}^{2}$ \cite{Lu-10prb,Shen-12book}, 
\begin{equation}
H=v\hbar k_{x}\sigma_{x}+v\hbar k_{y}\sigma_{y}+(mv^{2}-Bk^{2})\sigma_{z}.\label{eq:revised-Dirac}
\end{equation}
For a small $k$, the two models are almost identical if $Bk^{2}$
can be ignored. But for a larger $k$, the $Bk^{2}$ term in Eq. (\ref{eq:revised-Dirac}) will 
replace the linear term in determining the spin polarization.
Both the states of $\pm k_{x}$ are polarized along
the z-axis and the direction is determined by the sign of B. In this case 
the contradiction is removed. In the same condition,
the Hall conductivity has the same form by simply replacing $mv^{2}$
by $m(k_{c})$ in Eq. (\ref{eq:massive-Hall-conductivity}). In
a large $k_{c}$ limit, the Hall conductivity becomes 
$\sigma_{H}=\frac{e^{2}}{2h}\left[\mathrm{sgn}(m)+\mathrm{sgn}(B)\right]$ \cite{Lu-10prb}.
Thus the quadratic correction to the mass term brings an additional
one half to the Hall conductance. The Hall conductivity is zero
if $m$ and $B$ have the opposite sign, and equals to $\mathrm{sgn}(m)\frac{e^{2}}{h}$
if they have the same sign. Thus this massive Dirac fermions give an integer quantum Hall conductance, 
instead of one half.

\paragraph*{Realization on a lattice-}
The model in Eq. (\ref{eq:revised-Dirac}) is a prototype one in the study
of topological insulators, and can be realized on a lattice \cite{Shen-12book}. Making
a replacement, $k_{x}\rightarrow\frac{1}{a}\sin k_{x}a$, $k_{y}\rightarrow\frac{1}{a}\sin k_{y}a$
and $Bk^{2}\rightarrow\frac{4B}{a^{2}}\left(\sin^{2}\frac{k_{x}a}{2}+\sin^{2}\frac{k_{y}a}{2}\right)$
, one has 
\begin{equation}
H=\frac{v\hbar}{a}\left(\sin k_{x}a\sigma_{x}+\sin k_{y}a\sigma_{y}\right)+m_{eff}(k)\sigma_{z}
\end{equation}
with $m_{eff}(k)=mv^{2}-\frac{4B}{a^{2}}\left(\sin^{2}\frac{k_{x}a}{2}+\sin^{2}\frac{k_{y}a}{2}\right)$.
Note that $\sin^{2}\frac{k_{x}a}{2}$ instead of $\sin^{2}k_{x}a$, is used 
to avoid the next nearest hopping in the lattice model. Performing
the Fourier transformation, one has the tight-binding model just with
nearest hopping term on a lattice. For $m=0$, one has the massless Wilson fermions 
on a lattice, which was extensively studied in lattice gauge theory \cite{Wilson-75book,Rothe-05book}.

In the absence of the B term, there exist four fold degeneracy at
the points $(0,0)$, $(\frac{\pi}{a},0)$, $(0,\frac{\pi}{a})$, and
$(\frac{\pi}{a},\frac{\pi}{a})$. The model contains four massive
Dirac cones, leading to famous fermion doubling problem \cite{Nielsen-81prb}. The presence
of the B term removes the degeneracies and keep the minimal gap $2mv^{2}$
only at the point $(0,0)$. Thus in this way we can realize single
massive Dirac cone on a lattice. In this model, the Hall conductivity is always an
integer when the chemical potential is located within the band gap.
The Chern number $n_{c}=1$ if $0<mv^{2}a^{2}/B<4$, and $n_{c}=-1$
if $4<mv^{2}a^{2}/B<8$ and zero otherwise \cite{Shen-12book}. This is consistent with the Thouless-Khomoto-Nightingale-Njis
theorem \cite{Thouless-82prl}. Thus the model for massive Dirac fermions in Eq. (\ref{eq:massive-Dirac}) 
can not be realized on a lattice without modification. The half quantized Hall conductivity 
for massive Dirac fermions is an artifact of the continuous model, and is absent on a lattice. 

\paragraph*{Parity anomaly for massless or massive Dirac fermions?}
The k-dependent mass $m(k)$ in Eq. (\ref{eq:revised-Dirac}) reveals
the quantum anomaly in massless Dirac fermions instead of massive
Dirac fermions. When $m=0$, the energy gap closes at $k=0$ and the system is reduced to the massless Dirac
fermions although the B term is present. The Hall conductivity is a function of the
Fermi wave vector $k_{F}$,
\begin{equation}
\sigma_{H}=\frac{e^{2}}{2h}\left [\mathrm{sgn}(B)-\frac{Bk_{F}^{2}}{\sqrt{(v\hbar k_{F})^{2}+B^{2}k_{F}^{4}}} \right ],
\end{equation}
where the Fermi wave vector $k_{F}$ is determined by the chemical potential, $\mu_F^2=(v\hbar k_F)^2+(Bk_F)^2$. 
When the Fermi wave vector $k_{F}\rightarrow0$, that is, the chemical potential
is close to the crossing point, the conductivity is half quantized $\sigma_{H}=\frac{e^{2}}{2h}\mathrm{sgn}(B)$  
\cite{Fu-22npj,Shen-24Coshare}. In the case, 
the parity symmetry is restored near the Fermi level. When electrons 
move around the Fermi surface adiabactically, they acquire a Berry phase $\pi$, 
leading to the one half quantized Hall conductance \cite{Fu-22npj,Fu-25cp}.
This result actually reflects the parity anomaly for massless Dirac fermions. The B term acts as a regulator 
in the quantum field theory. The result is also valid for the model on a lattice. This result is opposite to Semenoff's proposal 
for massive Dirac fermions, but is consistent with the fact that the parity anomaly was 
first proposed for the massless Dirac fermions in quantum field theory, which
became more evident in Redlich's original work \cite{Redlich-84prl,Jackiw-84prd}.

There exists a singularity at the crossing point, which easily causes
some confusions. The Hall conductivity is a function of the mass as
well as the chemical potential. In the massive case of $m\neq0$,
if we take the chemical potential $\mu_{F}=0$ first, the Hall conductivity
is always an integer $\sigma_{H}=\frac{e^{2}}{2h}\left[\mathrm{sgn}(m)+\mathrm{sgn}(B)\right]$,
which persists even if $m\rightarrow 0$.
However, if we take the mass approach zero first, and then take the
chemical potential $\mu_{F}$ zero, the Hall conductivity is always
half quantized, $\sigma_{H}=\frac{e^{2}}{2h}\mathrm{sgn}(B)$. That means 
we have two unexchangeable limits at the crossing point, 
\begin{equation}
\lim_{m\rightarrow0}\lim_{\mu_{F}\rightarrow0}\sigma_{H}(m,\mu_{F})\neq\lim_{\mu_{F}\rightarrow0}\lim_{m\rightarrow0}\sigma_{H}(m,\mu_{F}).
\end{equation}
In Niemi and Semenoff approach (without the B-term) \cite{Niemi-83prl,Semenoff-84prl}, the chemical potential is equivalently
taken to be zero $\mu_{F}=0$, and then take $m\rightarrow0$. Thus they
had the half quantized Hal conductivity, $\sigma_{H}=\frac{e^{2}}{2h}\mathrm{sgn}(m)$.
However, if we take $m\rightarrow0$ first, then $\mu_{F}\rightarrow0$,
the Hall conductivity is actually zero, $\sigma_H=0$. In the quantum field theory,
the chemical potential is always exactly zero $\mu_{F}=0$ as all
states of the negative energy are fully filled in the vacuum. But in
the condensed matter physics, it is more reasonable to set the chemical potential zero after the zero mass 
due to the number and thermodynamic fluctuation of electrons in solids.

While the half quantized Hall conductivity appears only near the crossing
point in Eq. (\ref{eq:revised-Dirac}) at $m=0$, a further revision
of the mass term will give rise to the quantum plateau of the Hall conductivity
as a function of the chemical potential. The mass term is revised as $M(k)=\pm\Theta(-m(k))m(k)$ 
where $m(k)=mv^2-Bk^2$ and the step function $\Theta(x)=0$ for $x<0$ and +1 otherwise.
The mass correction appears only when $k>k_{c}=\sqrt{mv^{2}/B}$,
\begin{equation}
H=v\hbar k_{x}\sigma_{x}+v\hbar k_{y}\sigma_{y}\pm\Theta(-m(k))m(k)\sigma_{z}
\end{equation}
In this model the lower energy dispersions are linear and gapless for $k<k_{c}$. 
It does not break the parity symmetry and the time reversal symmetry. 
The Hall conductivity is $\sigma_{H}=\pm \frac{e^{2}}{2h}\mathrm{sgn}(B)$
when the chemical potential varies within the finite range of $\left|\mu_{F}\right|<v\hbar k_{c}$,
and decays to zero starting from $\mu_F>v\hbar k_{c}$. Again, the mass term $M(k)$
acts as a regulator in the quantum field theory. Here we have proposed an alternative 
approach to realize the parity anomaly for massless Dirac fermions, 
which is valid in either quantum field theory and condensed matter. This model 
can be realized on a lattice and derived from a three-dimensional continuous and lattice model as 
one for the surface states of topological insulator films,
and the sign is determined by the Zeeman field in the magnetically doped layer. The nonzero mass term 
is valid for the part of dispersions which enters the bulk \cite{Zou-23prb, Bai-24Post}.

\paragraph*{On Semenoff and Haldane proposals on a honeycomb lattice-}

Now we come to explain why Semenoff's proposal for quantum valley Hall effect 
\cite{Semenoff-84prl} fails 
but Haldane's proposal for quantum anomalous Hall effect \cite{Haldane-88prl} successes
as both of them used the same argument of massive Dirac fermions. The
Hall conductivity can be expressed as integral of the Berry curvature over
the whole Brillouin zone, which is equivalent to the Kubo formula \cite{Xiao-10rmp}. The honeycomb lattice hosts a pair of Dirac 
cones or valleys around the the point $\mathbf{K}$ and $\mathbf{K}'$. If we partition the Brillouin zone (BZ) into
two parts, and assume the dispersion has the masses at the two points are $M_{\pm}$,
we can write the Hall conductivity as 
\begin{equation}
\sigma_{H}=\frac{e^2}{h}\int_{BZ}\frac{d^2k}{2\pi}\Omega_z(k)=\sigma(M_{+})+\sigma(M_{-}).
\end{equation}
In Haldane's case, it is now known that the total Hall conductivity is quantized.
Thus his assumption becomes valid, $\sigma(M_{+})=\sigma(M_{-})=\mathrm{sgn}(M_{+})\frac{e^{2}}{2h}$, 
but it is still hard to partition the Brillouin zone because of unequal masses. Most important is that the emergence of the chiral
edge modes around the boundary makes the quantum anomalous Hall effect 
physically measurable for an insulating phases \cite{Hatsugai-93prl}.
As for Semenoff's case, the total Hall conductivity is zero,  $\sigma(M_{+})+\sigma(M_{-})=0$.
Although $\sigma(M_{+})-\sigma(M_{-})\approx \mathrm{sgn}(M_{+})\frac{e^{2}}{h}$,
it is an insulating state with a finite gap, and there does not
exist the extended edge states around the system boundary. Thus no extended state cross the Fermi level.
Now it it is questionable whether a quantity as an integral of the Berry curvature over part 
of the Brillouin zone is physically measurable or not. 
The interpretation $\sigma(M_{\pm})$ as the measurable Hall conductivity has no theoretical foundation. 
Also $\sigma(M_{\pm})$ are not exactly one half quantized as we show 
for a finite Brillouin zone, i.e., $k_c$ in Eq.(\ref{eq:massive-Hall-conductivity}) is finite.
In this case, the system is a trivial band insulator, and all electrons
are localized. It is impossible to have a response to a weak external
field. We have to point out that the effect is actually false. Some systems in which 
measured evidences for the quantum valley Hall effect
cannot be a true insulator, and must have partially populated 
bulk or edge states \cite{KazanTsev-24prl}.

\paragraph*{On the half quantized Hall conductivity of gapped surface states-}

The parity anomaly for massive Dirac fermions has been extensively
studied in the field of topological insulator \cite{Qi-11rmp,Tokura-19nrp,Chang-23rmp,Qi-08prb,Nomura-11prl}. 
The three-dimensional
topological insulator is surrounded by the gapless surface states. Magnetic
field or magnetic doping on the surface of topological insulator may
open an energy gap $V_{Z}$ for the surface states, forming single
massive Dirac fermions on one surface \cite{Chen-10science}. It is believed that this massive Dirac fermions
leads to the half quantized surface Hall effect \cite{Qi-08prb}. Consequently, it
is used to understand the topological magnetoeletric effect, the quantum anomalous Hall effect and axion
insulators. In fact the surface states exist within the bulk gap $\Delta$
of the topological insulator. Equivalently, the cutoff $k_c\approx \Delta/v\hbar$. 
Thus their contribution to the surface Hall conductivity
is approximately $\sigma_{surf}\approx\frac{e^{2}}{2h}\left[\mathrm{sgn}(V_{z})-\frac{V_{z}}{\Delta}\right]$,
which is obviously not quantized, and close to one half only when the ratio
$V_{Z}/\Delta$ is very tiny. However, the surface states is only
a small portion of the band in the Brillouin zone. Beyond the bulk
gap other part of the band also contribute a nonzero Hall conductance,
and effectively canceling the contribution from the surface states \cite{Zou-23prb}.
This cancellation is dedicated to the TKNN theorem \cite{Thouless-82prl}. So the so-called
half quantized Hall effect of the gapped surface states is baseless. 
Some topological magnetoelectric effects related to the gapped surface states
need to be reexamined carefully \cite{Li-10np}.

In the magnetically doped tri-layer topological insulator film, the
two outer layers are magnetically doped and magnetized, and the inner
layer is non-doped. When the magnetizations of two outer layers are
parallel, the quantum Hall conductance
is measured while when the magnetizations are antiparallel, the zero
Hall conductance is measured, which is usually named axion insulator 
\cite{Qi-08prb, Wang-15prb,Li-10np, Morimoto-15prb, Essin-09prl, 
Mogi-17nm, Xiao-18prl, Liu-20nm}. 
Layered $\mathrm{MnBi}_2\mathrm{Te}_4$ has similar structure and properties 
\cite{Li-18sciadv, Otrokov-19nature, Deng-20nature,Gao-21nature}.  
There are two surfaces in the topological insulator film, and each surface hosts massive Dirac fermions. 
Using the argument of parity anomaly for massive Dirac fermions, the quantum anomalous 
Hall effect is thought to be a result of addition of two half quantized Hall conductivities 
from the two surfaces, and the axion insulator is a result of their cancellation \cite{Tokura-19nrp,Chang-23rmp}. 
The accumulated layered Berry curvatures near the two surfaces 
are almost equal to one half, which is often used to support the argument \cite{Essin-09prl}.
However, detailed calculations of the band structures reveal that in the case of quantum anomalous Hall effect, two
bands are topologically trivial and two band are topologically nontrivial.
As for the case of axion insulator, all the four bands are topologically trivial \cite{Chen-24scp,Bai-24Post}. 
While the total Hall conductance is measurable and quantized, the difference of the top and bottom 
"layered Hall conductance" is not physically measurable just as quantum valley Hall conductance. 
The surface states on the two surface open an energy gap.
Unlike the quantum anomalous Hall effect in which there exist gapless chiral edge states around the boundary,
axion insulator is actually a trivial insulator. For a finite thick sample, the lateral surface states do exist, 
but are also gapped because of the finite size effect, 
which can be several meV for a sample with thickness up to a few hundred nanometers \cite{Zhou-23nc}. Thus,
the chiral edge current around the boundary must be absent if the chemical potential is within the gaps 
of top and bottom surface states and lateral surface states. 
Thus the axion insulator is very similar to quantum valley Hall effect, and is actually false, 
but the quantum anomalous Hall effect survives,
comparing with the Semenoff and Haldane propolsals on the honeycomb lattice.

The gate voltage may remove the degeneracy of the two massive Dirac fermions. 
If the chemical potential crosses one of the conduction or valence bands of the massive Dirac fermions,
the system becomes metallic and the Hall conductance is not quantized. In this case, the Hall conductance contributed 
by the massive Dirac fermions is not half quantized, and is a function of the chemical potential. 
The Hall conductance is attributted to the integral of non-zero Berry curvature from the massive Dirac fermions, 
and is not related to the parity anomaly.

\paragraph*{Coexistence of massless and massive Dirac fermions in magnetic topological insulators-}
Recently Mogi et al \cite{Mogi-22np} reported the experimental signature of the parity anomaly 
in a semi-magnetic topological insulator, and several other groups also reported the similar 
results \cite{Xiao-25xxx, Zhuo-25xxx, Muzaffar-25nano}.
Exprimental measurements show neither the Hall resistance nor the longitudinal resistance is quantized. 
Thus the system is not insulating but metallic because of the nonzero longitudinal conductivity. 
The system hosts massive Dirac fermions and massless Dirac fermions from the two surfaces of the 
topological insulator thin film, which coexist but the chemical 
potential is located witin the band gap of the massive Dirac fermions, and crosses the massless Dirac femions only. 
The chiral edge current is distributed in a power law decay from the edge into the bulk on the gapless surface, 
which is contributted collectively by the massless Dirac fermions. The half quantized Hall effect
is produced by the chiral edge currenrt \cite{Zou-22prb,Wang-24prb}. This is distinct from the quantum anomalous Hall effect, 
in which the extended edge states emerge and carry
an chiral edge current around the boundary in an exponential decay from the edge into the bulk. Although the 
massive Dirac fermions coexist, they do not contribute to the chiral edge current. It is worth emphasizing that, 
in the gapless Wilson fermions, the chiral edge current also exists and gives rise to half quantized Hall conductivity, 
in which there is no massive Dirac fermion \cite{Fu-22npj}. Thus the massive bands are not necessary for 
the half quantized Hall conductivity, although they coexist with the massless bands 
due to the time reversal symmetry breaking.

\paragraph*{Discussion and Conclusion}

Quantum anomaly refers to the failure of a symmetry present in a theory's
classical action to be a symmetry of any regulation of the fully quantum
theory. Strictly speaking, there is no parity anomaly in a solid or
on a lattice, as the lattice spacing in a solid crystal provides a
natural cut-off for the wave vector, and both the first Brillouin
zone and the band width are finite. In the case of the one-half quantum
Hall conductance, it is believed to be closely related with the physics
of parity anomaly. Near the Dirac crossing point or in the regime
of the gapless surface states, electrons respect the parity symmetry. 
However, the occupied electrons away from the Fermi surface break the symmetry, 
and contribute an exactly
half-quantized Hall conductance. In quantum field theory, the symmetry
breaking is induced by the regularization of the particles at higher
energy levels to remove the sign ambiguity, 
such as the Pauli-Villars regularization. In this sense,
the physics of parity anomaly occurs on a lattice system. As for massive 
Dirac fermions, the parity symmetry has been broken explicitly by the mass term, 
nonzero Hall conductance (if existing) is an explicit product of the broken symmetry. 
It is quite normal, NOT abnormal.

In short, the parity anomaly is absent for massive Dirac 
fermions on a lattice,  instead is an intrinsic property of 
a single massless Dirac cone of fermions on a lattice.

\paragraph{Acknowledgments-} 
The author would like to thank Qian Niu, Bo Fu, Kai-Zhi Bai, and Shi-Hao Bi for helpful discussions. 
This work was supported by the Research Grants Council, University Grants Committee, 
Hong Kong under Grants No. C7012-21G and No. 17301823; 
and Quantum Science Center of Guangdong-Hong Kong-Macao Greater Bay Area GDZX2301005.


\begin{thebibliography}{10}

\bibitem{Jackiw-84prd}R. Jackiw, Fractional Charge and Zero Modes for Planar 
Systems in a Magnetic Field, Phys. Rev. D 29, 2375 (1984).

\bibitem{Avareme-84npbz}L. Alvarez-Gaume and E. Witten, Gravitational Anomalies,
Nucl. Phys. B 234, 269 (1984).

\bibitem{Niemi-83prl}A. J. Niemi and G. W. Semenoff, Axial-Anomaly-Induced
Fermion Fractionization and Effective Gauge-Theory Actions in Odd-Dimensional
Space-Times, Phys. Rev. Lett. 51, 2077 (1983).

\bibitem{Redlich-84prl}A. N. Redlich, Gauge Noninvariance and parity Nonconservation 
of Three-Dimensional Fermions, Phys. Rev. Lett. 52, 18 (1984).

\bibitem{Semenoff-84prl}G. W. Semenoff, Condensed-Matter Simulation
of a Three-Dimensional Anomaly, Phys. Rev. Lett. 53, 2449 (1984).

\bibitem{Gorbachev-14science}R. V. Gorbachev, J. C. W. Song, G. L. Yu, A. V.
Kretinin, F. Withers, Y. Cao, A. Mishchenko, I. V. Grigorieva, 
K. S. Novoselov, L. S. Levitov, and A. K. Geim, 
Detecting topological currents in graphene superlattices, 
Science 346, 448 (2014).

\bibitem{Lensky-15prl}Y. D. Lensky, J. C. W. Song, P. Samutpraphoot, 
and L. S. Levitov, Topological Valley Currents in Gapped Dirac Materials, 
Phys. Rev. Lett. 114, 256601 (2015).

\bibitem{Ju-15nature}L. Ju, Zhiwen Shi, N. Nair, Y. Lv, C. Jin, J. Velasco Jr.,
C. Ojeda-Aristizabal, H. A. Bechtel, M. C. Martin, A. Zettl, J. Analytis, and F. Wang,
Topological valley transport at bilayer graphene domain walls, Nature 520, 650 (2015).

\bibitem{Sui-15natphys}M. Sui, G. Chen, L. Ma, W. Shan, D. Tian,
and K. F. Mak, Gate-tunable topological valley transport in bilayer
graphene, Nat. Phys. 11, 1027 (2015).

\bibitem{Shimazaki-15natphys}Y. Shimazaki, M. Yamamoto, I. V. Borzenets,
K. Watanabe, T. Taniguchi, and S. Tarucha, Generation and detection
of pure valley current by electrically induced Berry curvature in
bilayer graphene, Nat. Phys. 11, 1032 (2015).

\bibitem{Ren-16RPP}Y. Ren, Z. Qiao, and Q. Niu, Topological phases in two-dimensional
materials: a review, Rep. Prog. Phys. 79, 066501 (2016).

\bibitem{KazanTsev-24prl}A. Kazantsev, A. Millis, E. O'Neill, H. Sun, G. Vignale, and A. Principi,
Nonconservation of the Valley Density and Its Implicationsfor the Observation of the ValleyHall effect
Phys. rev. Lett. 132, 106301 (2024).
    
\bibitem{Haldane-88prl}F. D. M. Haldane, Model for a Quantum Hall
Effect without Landau Levels: Condensed-Matter Realization of the
``Parity Anomaly'', Phys. Rev. Lett. 61, 2015 (1988).

\bibitem{Chang-13science}C.-Z. Chang, J. Zhang, X. Feng, J. Shen,
Z. Zhang, M. Guo, K. Li, Y. Ou, P. Wei, L.-L. Wang, Z.-Q. Ji, Y. Feng,
S. Ji, X. Chen, J. Jia, X. Dai, Z. Fang, S.-C. Zhang, K. He, Y. Wang, L. Lu,
X.-C. Ma, and Q.-K. Xue, Experimental Observation of the Quantum Anomalous Hall Effect
in a Magnetic Topological Insulator, Science 340, 167 (2013).

\bibitem{Checkelsky-14natphys}J. G. Checkelsky, R. Yoshimi, A. Tsukazaki,
K. S. Takahashi, Y. Kozuka, J. Falson, M. Kawasaki, and Y. Tokura, Trajectory
of the anomalous Hall effect towards the quantized state in a ferromagnetic topological
insulator, Nat. Phys. 10, 731 (2014).

\bibitem{Kou-14prl}X. Kou, S.-T. Guo, Y. Fan, L. Pan, M. Lang, Y. Jiang, Q. Shao, T. Nie, 
K. Murata, J. Tang, Y. Wang, L. He, T. Kondo, S. Shin, K. L. Wang, and Y. Tokura, 
Scale-Invariant Quantum Anomalous Hall Effect in Magnetic Topological Insulators beyond the Two-Dimensional Limit,  
Phys. Rev. Lett. 113,137201 (2014).

\bibitem{Yu-10science}R. Yu, W. Zhang, H.-J. Zhang, S.-C. Zhang, X. Dai, and Z. Fang, Quantized Anomalous Hall Effect in Magnetic Topological Insulators,
Science 329, 61 (2010).

\bibitem{Chu-11prb}R.-L. Chu, J. Shi, and S.-Q. Shen, Surface Edge States
and Half-Quantized Hall Conductance in Topological Insulators, Phys. Rev. B 84, 085312 (2011).

\bibitem{Qiao-10prb}Z. Qiao, S. A. Yang, W. Feng, W.-K. Tse, J. Ding,
Y. Yao, J. Wang, and Q. Niu, Quantum Anomalous Hall Effect in Graphene
from Rashba and Exchange Effects, Phys. Rev. B 82, 161414(R) (2010).

\bibitem{Qi-11rmp}X. L. Qi and S. C. Zhang, Topological Insulators
and Superconductors, Rev. Mod. Phys. 83, 1057 (2011)

\bibitem{Tokura-19nrp}Y. Tokura, K. Yasuda, and A. Tsukazaki, Magnetic
Topological Insulators, Nat. Rev. Phys. 1, 126 (2019)

\bibitem{Chang-23rmp}C. Z. Chang, C. X. Liu, and A. H. MacDonald,
Quantum Anomalous Hall Effect, Rev. Mod. Phys. 95, 011002 (2023).
\bibitem{Xiao-10rmp}D. Xiao, M.-C. Chang, and Q. Niu, Berry Phase
Effects on Electronic Properties, Rev. Mod. Phys. 82, 1959 (2010).

\bibitem{Lu-10prb}H.-Z. Lu, W.-Y. Shan, W. Yao, Q. Niu, and S.-Q. Shen,
Massive Dirac Fermions and Spin Physics in an Ultrathin Film of
Topological Insulator, Phys. Rev. B 81, 115407 (2010).

\bibitem{Shen-12book}S.-Q. Shen, Topological Insulators: Dirac Equation
in Condensed Matters (Springer, Berlin, 2012).

\bibitem{Wilson-75book}K. G. Wilson, New phenomena in subnuclear physics. ed. A. Zichichi (New York, Plenum, 1975).

\bibitem{Rothe-05book}H. Rothe, Lattice Gauge Theories: An Introduction (World Scientific, Singapore, 2005).

\bibitem{Nielsen-81prb}H. B. Nielsen and M. Ninomiya, No Go Theorem
for Regularizing Chiral Fermions, Phys. Rev. B 185, 20 (1981).

\bibitem{Thouless-82prl}D. J. Thouless, M. Kohmoto, M. P. Nightingale, and M. den
Nijs, Quantized Hall Conductance in a Two-Dimensional Periodic Potential,
Phys. Rev. Lett. 49, 405 (1982).   

\bibitem{Fu-22npj}B. Fu, J.-Y. Zou, Z.-A. Hu, H.-W. Wang, and S.-Q.
Shen, Quantum anomalous semimetals, npj Quantum Mater. 7, 94 (2022).

\bibitem{Shen-24Coshare}S. Q. Shen, Half Quantized Hall Effect, Coshare
Science 2, 1 (2024)

\bibitem{Fu-25cp}B. Fu and S. Q. Shen, Z/2 topological invariant
and half quantized Hall effect, Communications Phys. 8, 2 (2025)

\bibitem{Zou-23prb} J.-Y. Zou, R. Chen, B. Fu, H.-W. Wang, Z.-A.
Hu, and S.-Q. Shen, Half-quantized Hall effect at the parity-invariant
Fermi surface Phys. Rev. B 107, 125153 (2023).

\bibitem{Bai-24Post}K. Z. Bai, B. Fu, and S.-Q. Shen, Dirac Fermions and 
Topological Phases in Magnetic Topological Insulator Films, SciPost Phys. 17, 146 (2024).

\bibitem{Hatsugai-93prl}Y. Hatsugai, Chern Number and Edge States
in the Integer Quantum Hall Effect, Phys. Rev. Lett. 71, 3697 (1993).
\bibitem{Li-10np}R. Li, J. Wang, X.-L. Qi, and S.-C. Zhang, Dynamical Axion Field
in Topological Magnetic Insulators, Nat. Phys. 6, 284 (2010).

\bibitem{Mogi-17nm}M. Mogi, M. Kawamura, R. Yoshimi, A. Tsukazaki,
K. S. Takahashi, M. Kawasaki, and Y. Tokura,
A magnetic heterostructure of topological insulators as a candidate for an axion insulator, Nat. Mater. 16, 516 (2017).

\bibitem{Xiao-18prl}D. Xiao, J. Jiang, J. H. Shin, W. B. Wang, F. Wang, 
Y. F. Zhao, C. X. Liu, W. D. Wu, M. H. W. Chan,  N. Samarth and C. Z. Chang,
Realization of the Axion Insulator State in Quantum Anomalous Hall
Sandwich Heterostructures, Phys. Rev. Lett. 120, 056801 (2018).

\bibitem{Liu-20nm}C. Liu, Y. C. Wang, H. Li, Y. Wu, 
Y. X. Li, J. H. Li, K. He, Y. Xu, J. S. Zhang, Y. Y. Wang,
Robust axion insulator and Chern insulator phases 
in a two-dimensional antiferromagnetic topological insulator,
Nat. Mater. 19, 522 (2020).

\bibitem{Qi-08prb}X.-L. Qi, T. L. Hughes, and S.-C. Zhang, Topological Field
Theory of Time-Reversal Invariant Insulators, Phys. Rev. B 78,195424 (2008).

\bibitem{Nomura-11prl}K. Nomura and N. Nagaosa, Surface-Quantized Anomalous Hall Current
and the Magnetoelectric Effect in Magnetically Disordered Topological Insulators,
Phys. Rev. Lett. 106, 166802 (2011).

\bibitem{Chen-10science}Y. L. Chen, J. H. Chu, J. G. Analytis, Z. K. Liu,
K. Igarashi, H.-H. Kuo, X. L. Qi, S. C. Zhang, I. R. Fisher, Z. Hussain,
and Z.-X. Shen, Massive Dirac Fermion on the Surface of a Magnetically
Doped Topological Insulator, Science 329, 659 (2010).

\bibitem{Wang-15prb}J. Wang, B. Lian, and S.-C. Zhang, Quantum topological magnetoelectric
effect and quantized Faraday and Kerr rotation in magnetic topological
insulators, Phys. Rev. B 92, 081107(R) (2015).

\bibitem{Essin-09prl}A. M. Essin, J. E. Moore, and D. Vanderbilt,
Magnetoelectric Polarizability and Axion Electrodynamics in Crystalline
Insulators, Phys. Rev. Lett. 102, 146805 (2009).

\bibitem{Morimoto-15prb}T. Morimoto, A. Furusaki, and N. Nagaosa, Topological
Magnetoelectric Effect in Thin Films of Topological Insulators, Phys. Rev. B 92, 085113 (2015).

\bibitem{Li-18sciadv}J. Li, Y. Li, S. Q. Du, Z. Wang, 
B. L. Gu, S. C. Zhang, K. He, W. H. Duan and Y. Xu,
Intrinsic magnetic topological insulators in van der Waals 
layered MnBi2Te4-family materials, Sci. Adv. 5, eaaw5685 (2020).

\bibitem{Otrokov-19nature}M. M. Otrokov, I. P. Rusinov, A. Y. Vyazovskaya, M. Amado, 
Y. M. Koroteev, J. Sánchez-Barriga, O. E. Tereshchenko, A. Ernst, P. M. Echenique, A. Arnau
, et al., Prediction and observation of an antiferromagnetic topological insulator, Nature 576, 416 (2019)  

\bibitem{Deng-20nature} Y. Deng, Y. Yu, M. Z. Shi, Z. Guo, Z. Xu, J. Wang, X. H. Chen, and Y. Zhang, 
Quantum anomalous Hall effect in intrinsic magnetic topological insulator MnBi2Te4, Nature 567, 94 (2020)    

\bibitem{Gao-21nature}A. Gao, Y. F. Liu, Chaowei Hu, J. X. Qiu et al., Nature 595, 521 (2021).

\bibitem{Chen-24scp}R. Chen, and S.-Q. Shen, On the half-quantized Hall conductance of massive surface 
electrons in magnetic topological insulator films, Sci. China Phys. Mech. Astron. 67, 267011 (2024).

\bibitem{Zhou-23nc}D.Y. Zhou, Z. J. Yan, Z. T. Sun, Y. F. Zhao, R. Zhang, R. Mei, H. Yi, K. Wang, M. H. W. Moses, C. X. Liu, K. T. Law and C. Z. Chang,
Axion insulator states in hundred-nanometer-thick magnetic topological insulator sandwich heterostructures, Nat. Commun. 14, 7596 (2023).   

\bibitem{Mogi-22np}M. Mogi, Y. Okamura, M. Kawamura, R. Yoshimi,
K. Yasuda, A. Tsukazaki, K. Takahashi, T. Morimoto, N. Nagaosa, M.
Kawasaki, et al., Experimental Signature of the Parity Anomaly in
a Semi-magnetic Topological Insulator, Nat. Phys. 18, 390 (2022).


\bibitem{Xiao-25xxx} D. Xiao, J. Hu, B. Wang, H. Zhou, T. Jia, Z. Sun, C. Liu, 
B. Zhang, D. Qian, T. Li, X. C. Xie, Y. Kong and C. Z. Chen,
Half-Quantized Layer Hall Effect as a Probe of Quantized Axion Field, arXiv: 2503.06021 (2025)

\bibitem{Zhuo-25xxx}D. Zhuo, B. Zhang, H. Zhou, H. Tay, X. Liu, Z. Xi, C. Z. Chen, and C. Z. Chang, Evidence for Half-Quantized Chiral Edge Current in a C = 1/2 Parity
Anomaly State, arXiv: 2509.15525 (2025)


\bibitem{Muzaffar-25nano}M. U. Muzaffar, K. Z. Bai, W. Qin, G. Cao, B. Fu, P. Cui, S. Q. Shen, and Z. Zhang,
Ferroelectrically Switchiable Half-Quantized Hall Effect, Nano Lett. 25, 7361 (2025).

\bibitem{Zou-22prb}J.-Y. Zou, B. Fu, H.-W. Wang, Z.-A. Hu, and S.-Q.
Shen, Half- quantized Hall effect and power law decay of edge current
distribution, Phys. Rev. B 105, L201106 (2022)

\bibitem{Wang-24prb} H.-W. Wang, B. Fu, and S.-Q. Shen, Signature
of parity anomaly: Crossover from one half to integer quantized Hall
conductance in a finite magnetic field, Phys. Rev. B 109, 075113 (2024)

\end{thebibliography}
\end{document}